\documentclass[ reprint, superscriptaddress, amsmath,amssymb,aps]{revtex4-2}

\usepackage{graphicx}
\usepackage{dcolumn}
\usepackage{bm}
\usepackage[mathlines]{lineno}
\begin{document}

\preprint{APS/123-QED}

\title{Observation of unexpected band splitting and magnetically-induced band structure reconstruction in TbTi$_3$Bi$_4$.
}

\author{Yevhen Kushnirenko}
\email[]{kushes16@gmail.com}
\affiliation{Division of Materials Science and Engineering, Ames National Laboratory, Ames, Iowa 50011, USA}
\affiliation{Department of Physics and Astronomy, Iowa State University, Ames, Iowa 50011, USA}

\author{Lin-Lin Wang}
\affiliation{Division of Materials Science and Engineering, Ames National Laboratory, Ames, Iowa 50011, USA}
\affiliation{Department of Physics and Astronomy, Iowa State University, Ames, Iowa 50011, USA}

\author{Xiaoyi Su}
\affiliation{Division of Materials Science and Engineering, Ames National Laboratory, Ames, Iowa 50011, USA}

\author{Benjamin Schrunk}
\affiliation{Division of Materials Science and Engineering, Ames National Laboratory, Ames, Iowa 50011, USA}

\author{Evan O'Leary}
\affiliation{Department of Physics and Astronomy, Iowa State University, Ames, Iowa 50011, USA}

\author{Andrew Eaton}
\affiliation{Division of Materials Science and Engineering, Ames National Laboratory, Ames, Iowa 50011, USA}
\affiliation{Department of Physics and Astronomy, Iowa State University, Ames, Iowa 50011, USA}

\author{P. C. Canfield}
\affiliation{Division of Materials Science and Engineering, Ames National Laboratory, Ames, Iowa 50011, USA}
\affiliation{Department of Physics and Astronomy, Iowa State University, Ames, Iowa 50011, USA}

\author{Adam Kaminski}
\email[]{kaminski@ameslab.gov}
\affiliation{Division of Materials Science and Engineering, Ames National Laboratory, Ames, Iowa 50011, USA}
\affiliation{Department of Physics and Astronomy, Iowa State University, Ames, Iowa 50011, USA}

\begin{abstract}
The magnetic Kagome materials are a promising platform to study the interplay between magnetism, topology, and correlated electronic phenomena. Among these materials, the RTi$_3$Bi$_4$ family received a great deal of attention recently because of its chemical versatility and wide range of magnetic properties. Here, we use angle-resolved photoemission spectroscopy measurements and density functional theory calculations to investigate the electronic structure of TbTi$_3$Bi$_4$ in paramagnetic and antiferromagnetic phases.  Our experimental results show the presence of unidirectional band splitting of unknown nature in both phases. In addition, we observed a complex reconstruction of the band structure in the antiferromagnetic phase. Some aspects of this reconstruction are consistent with effects of additional periodicity introduced by the magnetic ordering vector, while the nature of several other features remains unknown.
\end{abstract}

\maketitle

\section{Introduction}

The Kagome materials have attracted substantial research interest
due to their rich electronic structure and properties \cite{wang2024topological, negi2025magnetic}.  The Kagome motif renders a two-dimensional honeycomb network of corner-sharing triangles. Such a structure possesses inherent geometric frustration that, in the simplest case, results in an electronic structure with flat bands, van Hove singularities (VHs), and Dirac cones \cite{bergman2008band, guo2009topological, tang2011high, kiesel2012sublattice, wang2013competing, kiesel2013unconventional, meier2020flat, park2021electronic, neupert2022charge, yin2022topological}. These
materials exhibit diverse exotic quantum phenomena such as charge density waves \cite{jiang2021unconventional, arachchige2022charge, zhang2022destabilization, hu2023optical}, superconductivity \cite{ortiz2019new, ortiz2020cs, yin2021superconductivity, jiang2023kagome}, anomalous Hall effect \cite{nakatsuji2015large, nayak2016large, liu2018giant, yang2020giant, chen2021large, yu2021concurrence}, and topological Hall effect \cite{ma2025anisotropic, wang2021field, ghimire2020competing, dhakal2021anisotropically, kabir2022unusual, wang2021field}.

The crystal structure of RTi$_3$Bi$_4$, where R is a rare-earth element, materials (134 family) is shown in Fig.~1a. In these materials, the Kagome sublattice is built by non-magnetic Ti atoms, while quasi-one-dimensional zig-zag chains of rare-earth atoms are responsible for magnetism. The possibility of tuning magnetism in these materials from nonmagnetic to ferromagnetic (FM) and antiferromagnetic (AFM) by changing rare-earth atoms \cite{ortiz2023evolution, hu2024magnetic, ortiz2024intricate, chen2024tunable} makes this family of Kagome materials a fertile ground to study the interplay of magnetism and topologically nontrivial electronic structure that results from the Kagome motif. 
TbTi$_3$Bi$_4$ was shown to have a complex magnetic landscape. An AFM transition at T~=~20.4~K was observed in the magnetization, resistivity, and specific heat measurements \cite{ortiz2024intricate, guo20241}. Besides that, the magnetization measurements showed 1/3 and 2/3 magnetization plateaus and multiple metamagnetic transitions when an external magnetic field is applied. TbTi$_3$Bi$_4$ demonstrates an exceptionally large anomalous Hall effect, with a Hall coefficient of 6.2$\times10^5 \Omega^{-1}cm^{-1}$ \cite{cheng2024giant}.


In this work, we present the results of our density functional theory (DFT) and angle-resolved photoemission spectroscopy (ARPES) study of the electronic structure of TbTi$_3$Bi$_4$ in the paramagnetic (PM) and AFM state. We found an unconventional, unidirectional splitting of the bands in the PM state and complex reconstruction of the electronic structure in the AFM state. Even though the band folding vector agrees with the periodicity and direction of the magnetic ordering, this reconstruction includes the development of additional states that break mirror symmetry and cannot be explained by such band folding.

\begin{figure*}[t]
    \includegraphics[width=0.7\linewidth]{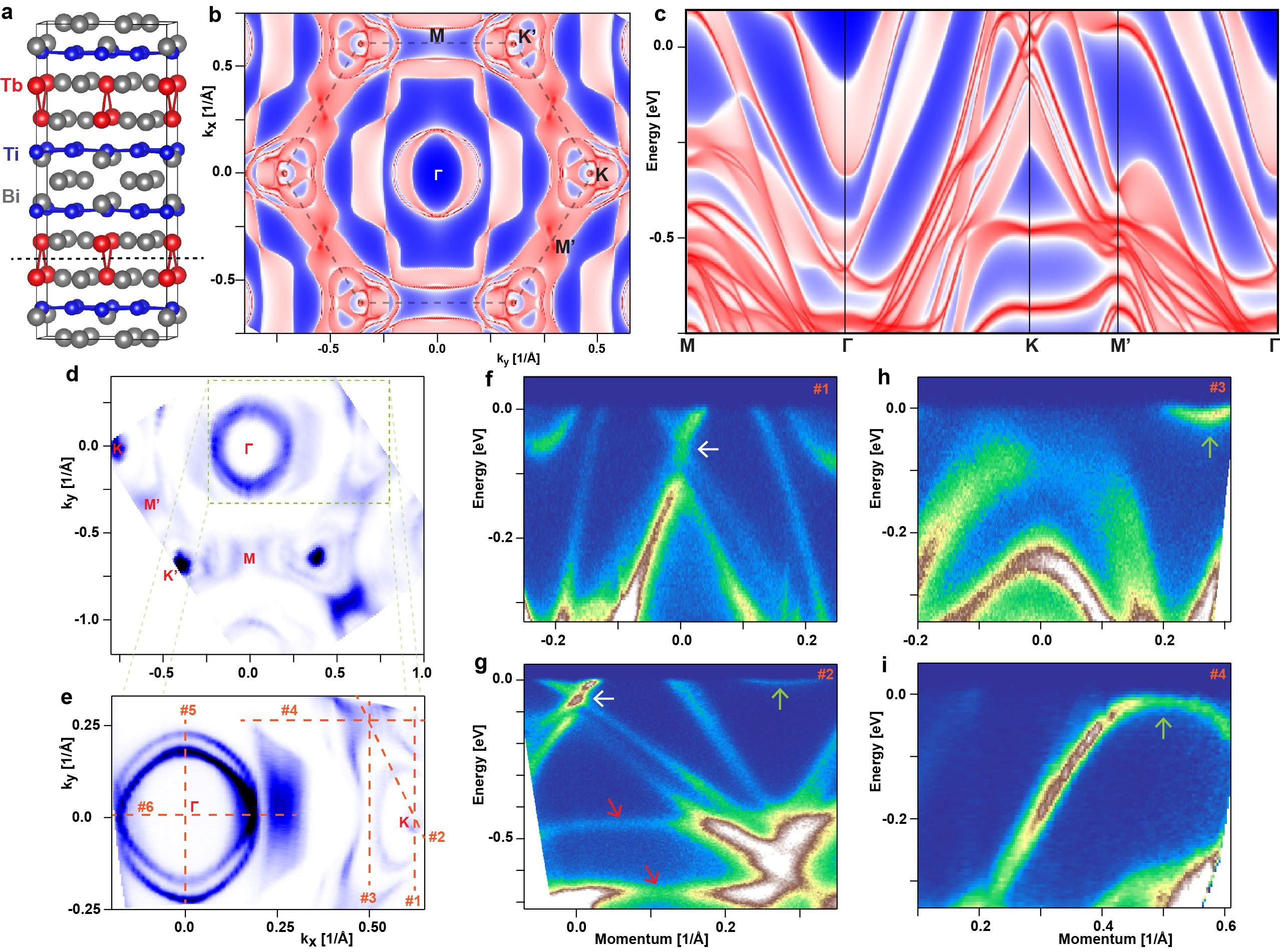}%
    \caption{Electronic structure (paramagnetic state).
    (a) Crystal structure of TbTi$_3$Bi$_4$. The dashed line depicts the termination plane.
    (b) and (c) DFT calculated Fermi surface and band dispersions along high-symmetry directions of the (001) surface.
    (d) Fermi surface map measured using 21.2~eV light.
    (e) Fermi surface map measured at T~=~28~K using 6.7~eV light. It represents the area marked with a green box in (d).
    (f-i) Band dispersions along cuts marked in (e) with dashed lines.
    }
\end{figure*}

\section{Experimental details}
Single crystals of TbTi$_3$Bi$_4$ were synthesized out of excess Bi flux \cite{guo20241} by using the high-temperature solution growth method \cite{canfield2016use}. Small pieces of Tb (99.97\%), Ti (99.97\%), and Bi (99.99\%), with a starting composition of Tb$_1$Ti$_1$Bi$_{20}$ were weighed and placed into the bottom of an alumina Canfield crucible set (CCS) \cite{canfield2019new, CrucibleSets}.
The packed CCS was flame-sealed into evacuated fused silica ampules that were backfilled with $\approx$~1/6 atm Ar gas.
Using a box furnace, the ampoules were heated to 1180~\textdegree C over 6 hours, dwelled at 1180~\textdegree C for 12 hours, cooled to 1000~\textdegree C, and then further cooled to 500~\textdegree C over 100 hours.
After that, the mixture was annealed at 500~\textdegree C for over 8 days. The excess liquid was then decanted by inverting the ampoules into a specially designed centrifuge with a metal rotor and cups \cite{canfield2019new}. The ampoules were opened at room temperature, and hexagonal crystals of TbTi$_3$Bi$_4$ were obtained and characterized by single-crystal X-ray diffraction (see Appendix E).

The band structures of TbTi$_3$Bi$_4$ have been calculated in DFT \cite{Hohenberg1964Inhomogeneous, Kohn1965Self} with spin-orbit coupling (SOC) in the PBE \cite{PBE} exchange-correlation functional using a plane-wave basis set and projector augmented wave method \cite{Blochl94PRB}, as implemented in the Vienna Ab-initio Simulation Package (VASP) \cite{Kresse96CMS, Kresse96PRB}. In the DFT calculations, we used a kinetic energy cutoff of 178~eV, $\Gamma$-centered Monkhorst-Pack \cite{Monkhorst1976Special} (8$\times$8$\times$3) k-point mesh and a Gaussian smearing of 0.05 eV with the experimental lattice constants of a~=~5.8681, b~=~10.3473, c~=~14.785~\AA. For the non-magnetic phase, the strongly localized Tb 4f orbitals have been treated as core electrons. Using maximally localized Wannier functions \cite{Marzari1997Maximally, Souza2001Maximally}, tight-binding models were constructed to reproduce closely the band structure including SOC within E$_F \pm$1eV with Tb s-d, Ti s-d, and Bi p orbitals. The surface spectral function and 2D Fermi surface (FS) were calculated with the surface Green’s function methods \cite{Sancho1984Quick, Sancho1985Highly} as implemented in WannierTools \cite{wu2018wanniertools}.

Most of the ARPES data were collected using vacuum ultraviolet (VUV) laser ARPES spectrometer that consists of a Scienta DA30 electron analyzer, tunable picosecond Ti:Sapphire oscillator paired with fourth-harmonic generator \cite{jiang2014tunable}. Data were collected with 6.2-6.9~eV photon energy and linear vertical, circular left, and circular right polarization. Angular resolution was set at $\sim$ 0.1$^{\circ}$ and 1$^{\circ}$, along and perpendicular to the direction of the analyzer slit, respectively, and the energy resolution was set at 2~meV. The diameter of the photon beam on the sample was $\sim 15\,\mu$m. The measurements at photon energy of 21.2~eV were carried out using R8000 analyzer and GammaData helium discharge lamp with custom focusing optics. The diameter of the photon beam on the sample was $\sim1$~mm. Samples were cleaved \textit{in-situ} along (001) plane, usually producing very flat, mirror-like surfaces. The measurements were performed at a base pressure lower than 2$\times$10$^{-11}$ Torr.

\section{Results and Discussion}

In Fig.~1a-c, we show the results of our semi-infinite surface calculations for nonmagnetic TbTi$_3$Bi$_4$. The calculations were performed for Tb-termination since RTi$_3$Bi$_4$ crystals are known to exfoliate between these two layers \cite{hu2024magnetic, cheng2024giant}. The presence of quasi-1D chains of rare-earth atoms and the deformation of the kagome layer \cite{ortiz2024intricate, sakhya2024diverse} break the rotational symmetry in these materials, which is clearly manifested in the calculated FS. Hence, strictly speaking, the BZ in this material is orthorhombic. However, for consistency with other kagome materials, we use quasi-hexagonal notation. The calculations predict the electronic structure similar to those of other members of the RTi$_3$Bi$_4$ family \cite{jiang2023direct, sakhya2024diverse, park2024spin, hu2024magnetic, mondal2023observation, zheng2024anisotropic, sakhya2025diverse}. Particularly, it has VHs at $M$ and $M'-$points, a Dirac point near $K$ and $K'-$points, and flat bands (see Fig.~1c and Fig.~8). The round pocket around the $\Gamma-$point is mostly formed by Bi~$6p$ states, and the outer pocket around the $\Gamma-$point is mostly formed by Tb~$5d$ states.

\begin{figure*}[t]
    \includegraphics[width=0.75\linewidth]{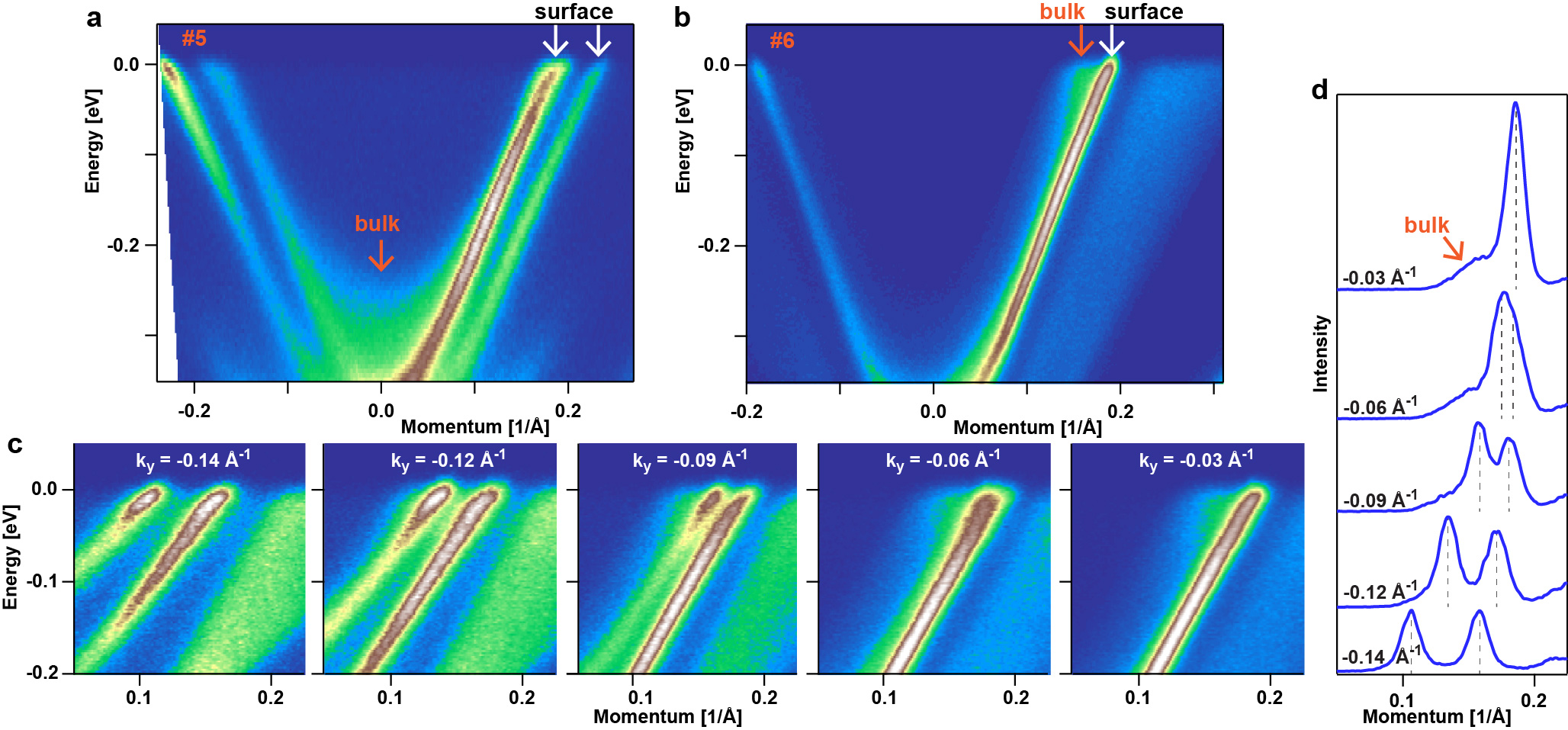}%
    \caption{Splitting of the electron-like band.
    (a-b) Spectra measured at T~=~28~K along $\Gamma-M$ and $\Gamma-K$ directions, respectively, (cuts \#5 and 6 in Fig.~1e).
    (c) Spectra measuredat T~=~28~K parallel to the $\Gamma-K$ direction at different $k_y$s plots.
    (d) Fermi level MDCs for spectra in (c).
    }
\end{figure*}

In Fig.~1d and e, we demonstrate two Fermi surface maps measured using He-lamp and laser, respectively. Both maps demonstrate two sharp circular pockets around the $\Gamma-$point, a broad arc-like feature along the $\Gamma-K$ direction, a small round pocket, and two bigger triangular pockets around the $K(K')-$point. The spectra measured through $K-$point along $K-M'$ direction (Fig.~1f) and the direction perpendicular to $\Gamma-K$ (Fig.~1g) demonstrates the presence of a Dirac-cone at E$_B$~=~60~meV (marked with white arrows). Also in Fig.~1g, one can see two flat bands at E$_B$~=~480 and 697~meV (marked with red arrows). Two orthogonal cuts through $M'-$point in Fig.~1h and i demonstrate a band that has electron-like character along one direction and hole-like character along another direction (marked with cyan arrows), indicating the presence of a saddle-point VHs near the Fermi level at the $M'-$point. 
The observed Dirac points, flat bands, and VHs are in agreement with our DFT calculations and earlier ARPES studies of TbTi$_3$Bi$_4$ \cite{cheng2024giant, zhang2024observation}.

The main difference between the experimental results and the DFT calculations is the presence of two sharp circular pockets around the $\Gamma-$point in the experiment. 
As can be seen from the spectrum measured along the $\Gamma-M$ direction (Fig.~2a), the splitting between the two electron-like bands that form these pockets is present across the whole energy range. These bands are also sharp across the whole energy range, which can indicate their surface state (SS) nature. At first glance, this splitting can look similar to the splitting of the electron-like pocket in the AV$_3$Sb$_5$ kagome family, which was explained by splitting between bulk and surface states by effects of surface self-doping \cite{kato2023surface}. However, in contrast to concentric circles in AV$_3$Sb$_5$ materials, in TbTi$_3$Bi$_4$, we observe two seemingly identical circles shifted along $k_y$. This can also be seen in the spectrum measured along the to $\Gamma-K$ direction (Fig.~2b), where two SS dispersions merge, so only one sharp feature can be observed. To show this better, we also analyzed spectra measured parallel to $\Gamma-K$ (Fig.~2c). The splitting is largest for k$_y$~=~-0.14~\AA$^{-1}$ while in cuts closer to the high symmetry direction,  two bands get closer. At k$_y$~=~-0.06~\AA$^{-1}$, two bands visually merge. However, the resulting feature looks broader, and the fitting of the Fermi level momentum distribution curves (MDC) in Fig.~2d still shows splitting of around 0.01~\AA$^{-1}$. In the spectrum measured at k$_y$~=~-~0.03~\AA$^{-1}$, no splitting can be detected, indicating that the bands that are split along $\Gamma-M$ direction, indeed,  merge along $\Gamma-K$ direction.

\begin{figure}[b]
    \includegraphics[width=1.0\linewidth]{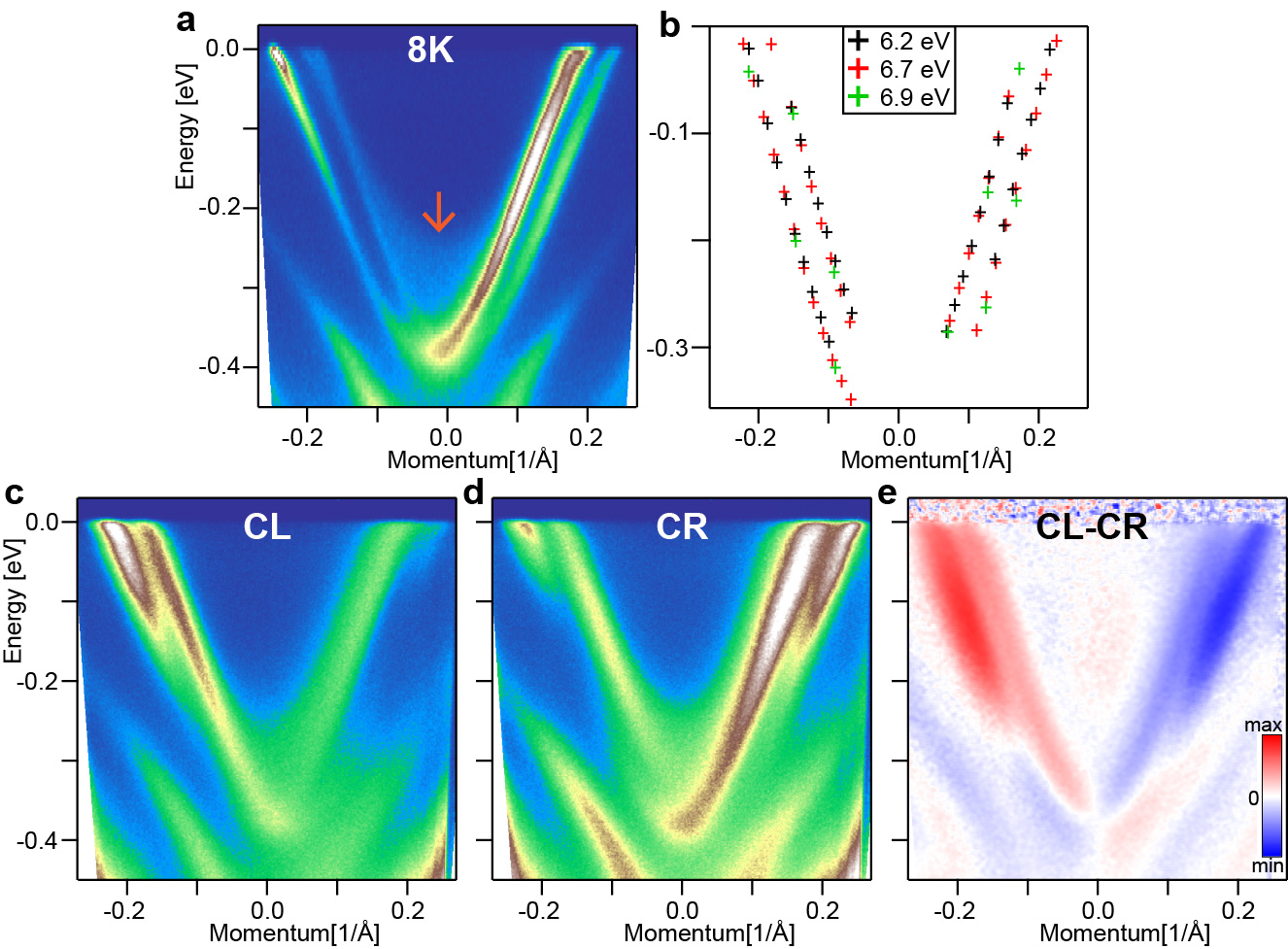}%
    \caption{Splitting of the electron-like band: k$_z$ dispersion and circular dichroism.  
    (a) Band dispersions in the magnetically ordered state (T~=~8~K) along $\Gamma-M$ direction measured using linear verticallly polarized light.
    (b) Comparison of the surface state band shape expected from datasets measured using different photon energy.
    (c), (d) Band dispersions along $\Gamma-M$ direction measured at T~=~7~K using CL and CR polarized light, respectively; and (e) is the corresponding circular dichroism plot.
    }
\end{figure}

\begin{figure*}[t]
    \includegraphics[width=0.85\linewidth]{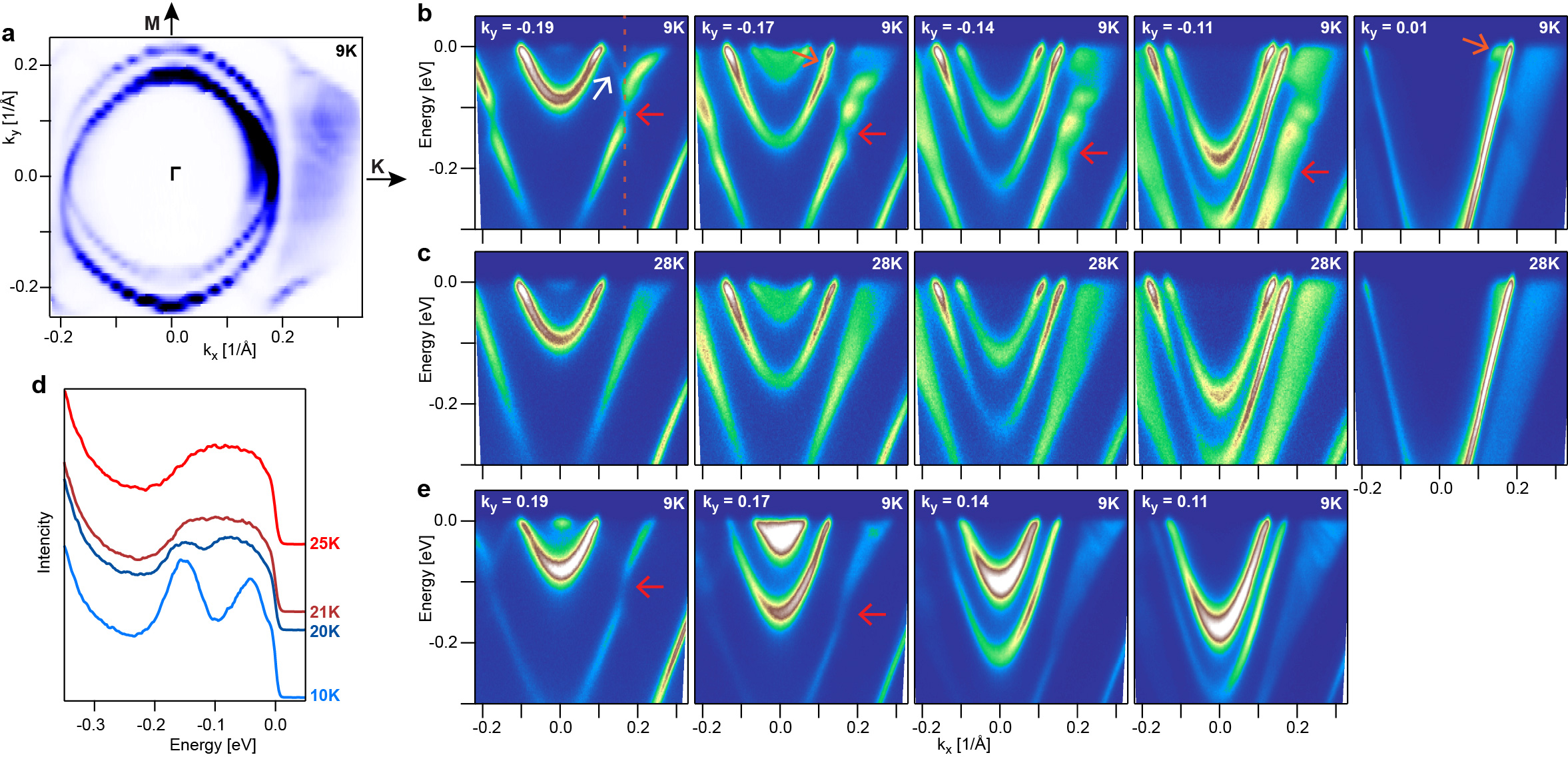}%
    \caption{Band structure reconstruction in the magnetically ordered state: band folding.
    (a) Fermi surface map measured at T~=~9~K. 
    (b) Corresponding spectra measured parallel to the $\Gamma-K$ direction at different $k_y$s.
    (c) The same as (b) but for paramagnetic state (T~=~28~K). 
    (e) The same as (b) but for positive $k_y$ values.
    (d) EDCs taken along the direction marked in (b) by a dashed line from spectra measured at different temperatures. EDCs were integrated over 0.08~\AA$^{-1}$ range. For more data, see Appendix B.
    }
\end{figure*}

The transition to the magnetically ordered state does not affect these SS significantly. The spectrum measured at T~=~8~K along the $\Gamma-M$ direction (Fig.~3a) has the same two SS bands as Fig.~2a. This makes this splitting different from the splitting in NdTi$_3$Bi$_4$, which can be observed in the FM state and disappears in the PM state \cite{hu2024magnetic}. At first glance, the parabolic bands that are split in momentum resemble a case of Rashba splitting \cite{rashba1959symmetry}. Unidirectional Rashba splitting was reported in a single-layer WS$_{1.4}$Se$_{0.6}$ alloy, where it is a result of giant spontaneous in-plane electrical polarization \cite{zribi2022unidirectional}. However, splitting in TbTi$_3$Bi$_4$ cannot be explained this way since this material is metal and thus can not be ferroelectric. For further analysis, we measured circular dichroism. The spectra in Fig.~3c and d were measured along the $\Gamma-M$ direction using circular left (CL) and right (CR) polarized light, respectively. In Fig.~1e, we show the corresponding circular dichroism plot, which is the difference in intensity between the spectra measured with CL and CR polarized light divided by their sum. 
Both bands in the left half of the spectrum get suppressed when polarization is switched from CL to CR, while bands in the right half of the spectrum gain intensity. The redistribution of the intensity from left to right of the spectrum upon changing polarization can be explained by geometric effects \cite{venus1993magnetic}. The identical behavior of the inner and outer bands indicates that either they have the same spin polarization, or, more likely in this case, they are not spin-polarized. This result again contradicts the Rashba scenario, in which the inner and outer bands are expected to have opposite spin polarizations.

To show that these states are indeed SS, we extracted the shape of these bands from three data sets measured using different photon energies that correspond to different $k_z$ values by fitting energy distribution curves (EDC). The results are shown in Fig.~3b demonstrate that band dispersion, energy position and shape of the bands are the same for all three photon energies, confirming that these states are indeed SS. As we already mentioned, the calculations predict the presence of a bulk electron-like band at the $\Gamma-$point. The signature of this band can also be identified in our experimental data. Particularly, besides two sharp parabolas, one can see some very broad features in Fig~2a and Fig~3a below E$_B$~=~200~meV (marked with an orange arrow). One can also see a relatively broad shoulder on the left of the SS band in the $\Gamma-K$ spectrum (Fig.~2b) and the last three spectra in Fig.~2c. Therefore, we associate these features with the bulk Bi~$6p$ band.

Another broad feature in Fig.~2b, which is located to the right of the SS, is formed by Tb~$5d$ bulk states and gives rise to the broad arc-like feature in the FS. We should note that this feature is not an actual arc. In fact, the DFT calculations (Fig.~1b) show that it is a part of a closed pocket. In our ARPES data, the electron-like feature that forms this pocket can still be observed at $|$k$_y| >$~0.2~\AA$^{-1}$; however, they are strongly suppressed, especially near the Fermi level (see Appendix A).

\begin{figure*}[t]
    \includegraphics[width=0.80\linewidth]{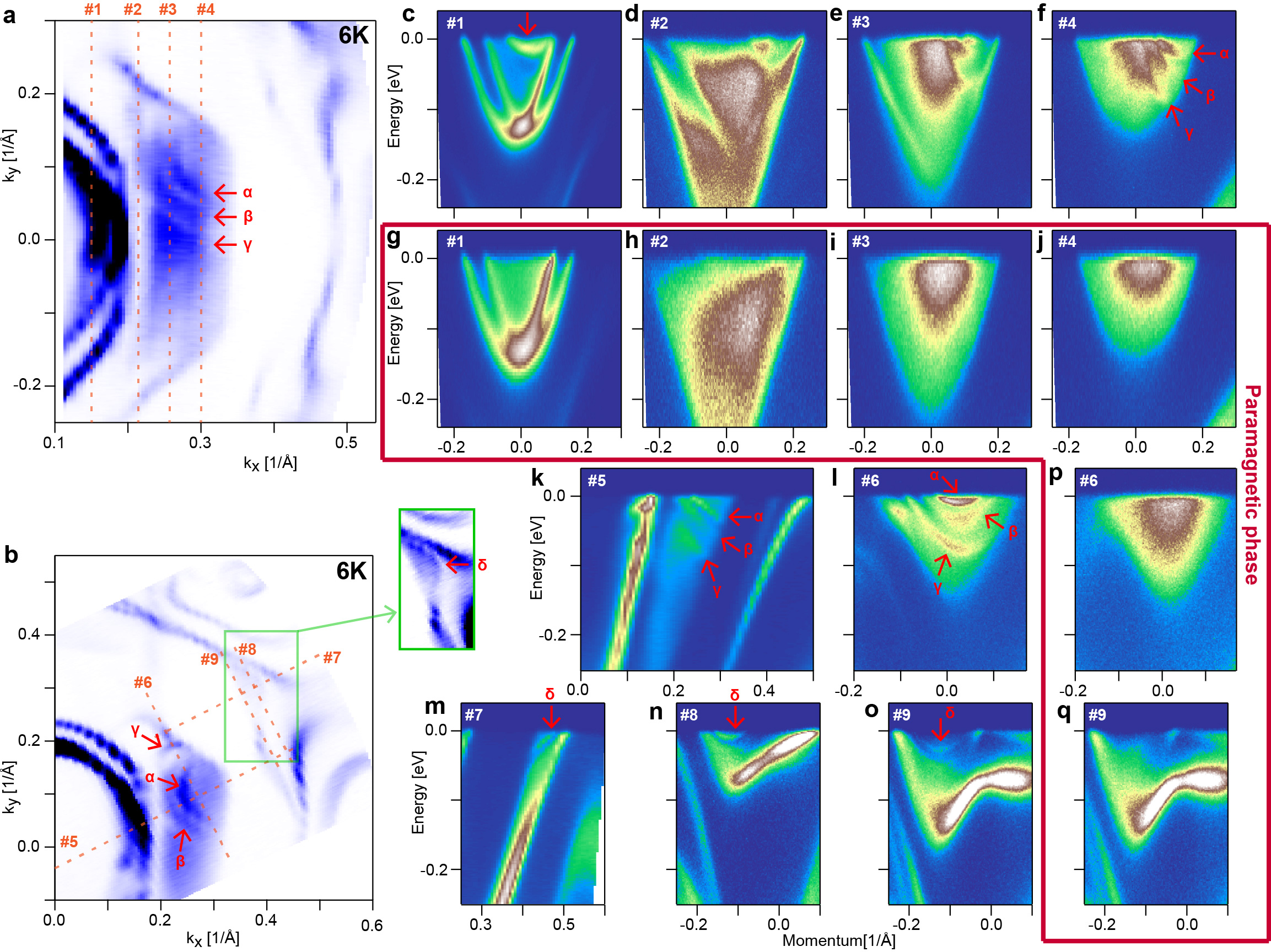}%
    \caption{Band structure reconstruction in the magnetically ordered state: surface resonance states.
    (a, b) Fermi surface maps measured at T~=~6~K from the sample rotated azimuthally from the position in Fig.~3a by 90 and 60 degrees, respectively. The resulting data was rotated back to match the map's orientation in Fig.~3a. The inset in (b) shows the enhanced intensity part of the FS marked by the green box.
    (c-f) band dispersions along cuts marked in (a).
    (g-j) the same as (b-d) but at T~=~28~K.
    (k-o) band dispersions along cuts marked in (b).
    (p) and (q) the same as (l) and (o) but at T~=~28~K. 
    }
\end{figure*}

Although we saw no changes in bands along the $\Gamma-M$ direction upon magnetic ordering, we observed significant changes in other parts of the band structure. The FS map measured at T~=~9~K is shown in Fig.~4a. Here, we can see the development of several sharp features over the broad Tb~$5d$ pocket. For further analysis of the changes, we plot spectra measured parallel to the $\Gamma-K$ direction at different $k_y$s in Fig.~4b and e and compare them with corresponding spectra measured in the PM state (Fig.~4c). In the spectra measured at $|$k$_y|$~=~0.19~\AA$^{-1}$, we observe the gap opening in both left and right branches of the Tb~$5d$ band and the emergence of a folded band (marked with a white arrow) upon magnetic ordering. The comparison of EDCs extracted at k$_x$~=~0.172~\AA$^{-1}$ from spectra measured at different temperatures (Fig.~4d) shows that the gap can be observed up to T~=~20~K. At T~=~21~K, the gap disappears, and upon further temperature increases, the shape of EDC does not change significantly. This result agrees with the early reported transition temperature of 20.4~K \cite{guo20241, ortiz2024intricate}. While shifting towards $\Gamma-K$ (cuts at smaller $|$k$_y|$), the gap in Tb~$5d$ band moves to higher binding energies (for more data, see Fig.~6). Further analysis has shown that the gap is located at k$_x$~=~$\pm$0.172~\AA$^{-1}$ for all spectra. This means that the gap is situated along two parallel lines connected with vector $q$~=~0.344~\AA$^{-1} \approx 2\pi/3a$. This agrees with $3a$ periodicity of magnetic order reported earlier \cite{guo20241, cheng2024giant}. So, the observed gap is a result of the band back folding and subsequent hybridization of one branch of the Tb~$5d$ band with another branch folded by a vector of $2\pi/3a$ along k$_x$. This band folding also affects Bi~$6p$ surface states 
(see orange arrows in the k$_y~=~-$0.17~\AA$^{-1}$spectrum). 
The observed band folding by $q$~=~$2\pi/3a$ agrees with earlier ARPES studies\cite{cheng2024giant, zhang2024observation}.

Besides the already described hybridization gap in the Tb~$5d$ band, the spectra in Fig.~4b and e demonstrate the development of other previously not reported features within this band upon the magnetic transition. Interestingly, these changes are not identical for positive and negative k$_y$ regions. Hypothetically, such asymmetry in ARPES data can be caused by matrix elements. To rule out this scenario, we performed additional measurements on samples rotated azimuthally by 90 degrees (Fig.~5a, c-j) and 60 degrees (Fig.~5b, k-r). For easier visual comparison, we rotated the resulting FS maps to match the FS orientation in Fig.~4a. All three FS maps (Fig.~5a,b and Fig.~4a) look alike, with several arc-like features in the positive k$_y$ region that are absent in the negative k$_y$ region. The cuts \#2-4 also demonstrate a complex asymmetric reconstruction of the Tb~$5d$ states upon magnetic ordering. The cuts \#4-6 demonstrate that the arc-like features on the FS are formed by three sharp electron-like dispersions. These dispersions are sharp and exist only in the areas where the bulk states are located. Thus, we can conclude that these states are surface resonance states \cite{zangwill1988physics}. We have also observed the development of a surface resonance in the bulk Bi~$6p$ states: it appears as an electron-like dispersion near the Fermi level in Fig.~5c and a small feature in Fig.~4b (k$_y$~=~0.01~\AA$^{-1}$). One more surface resonance was observed near the $M-$point. As can be seen from cuts \#7-9 (Fig.~5m-o) and the inset of Fig.~5b, it has an electron-like shape and forms an arc-like feature on the FS. In the PM state, these bands disappear. 

This is not the first instance when the SS Fermi arc-like features appear upon AFM transition \cite{schrunk2022emergence, kushnirenko2022rare}. However, what makes them unusual in the current case is their asymmetry. Such asymmetric states cannot be a result of simple band folding with a vector of $2\pi/3a$ along k$_x$. Most likely, these asymmetric states are the result of a surface reconstruction in the magnetically ordered state. The $\sqrt{3}\times\sqrt{3}$ surface reconstruction induced by magnetic order was previously reported in TbTi$_3$Bi$_4$ by STM measurements \cite{cheng2024giant}. However, such reconstruction would not break the mirror symmetry and thus cannot result in such states. So, the actual surface reconstruction in this material should be more complex.

\section{CONCLUSIONS}
We analyzed the electronic structure of TbTi$_3$Bi$_4$ in both paramagnetic and antiferromagnetic states using ARPES and DFT calculations. In addition to the presence of flat bands, van Hove singularities, and Dirac cones typical for the 134 family, we observed unconventional, unidirectional momentum band splitting that is not Rashba splitting \cite{rashba1959symmetry, zribi2022unidirectional}. In the antiferromagnetic states, we observed band folding with a vector that agreed with 3a periodicity of the magnetic order. Also, we observed the development of additional states that are asymmetric and cannot be explained by simple band-folding.

\begin{acknowledgments}
This work was supported by the U.S. Department of Energy, Office of Basic Energy Sciences, Division of Materials Science and Engineering. Ames National Laboratory is operated for the U.S. Department of Energy by Iowa State University under Contract No. DE-AC02-07CH11358.
\end{acknowledgments}

\appendix
\section{Band structure in the magnetically ordered state (additional data).}
The equal-energy maps obtained at E$_B$=~140, 210, and 280~meV show that Tb~$5d$ band forms a closed pocket around the $\Gamma$-point (Fig.~6a). At E$_B$=~140~meV, the intensity of some parts of that pocket gets suppressed, and at the FL, those parts fade entirely. We associate this suppression with the matrix elements in the ARPES experiment and think that Tb~$5d$ band forms a closed pocket at all energies, as predicted by the DFT calculations.
\begin{figure*}[b]
    \includegraphics[width=0.8\linewidth]{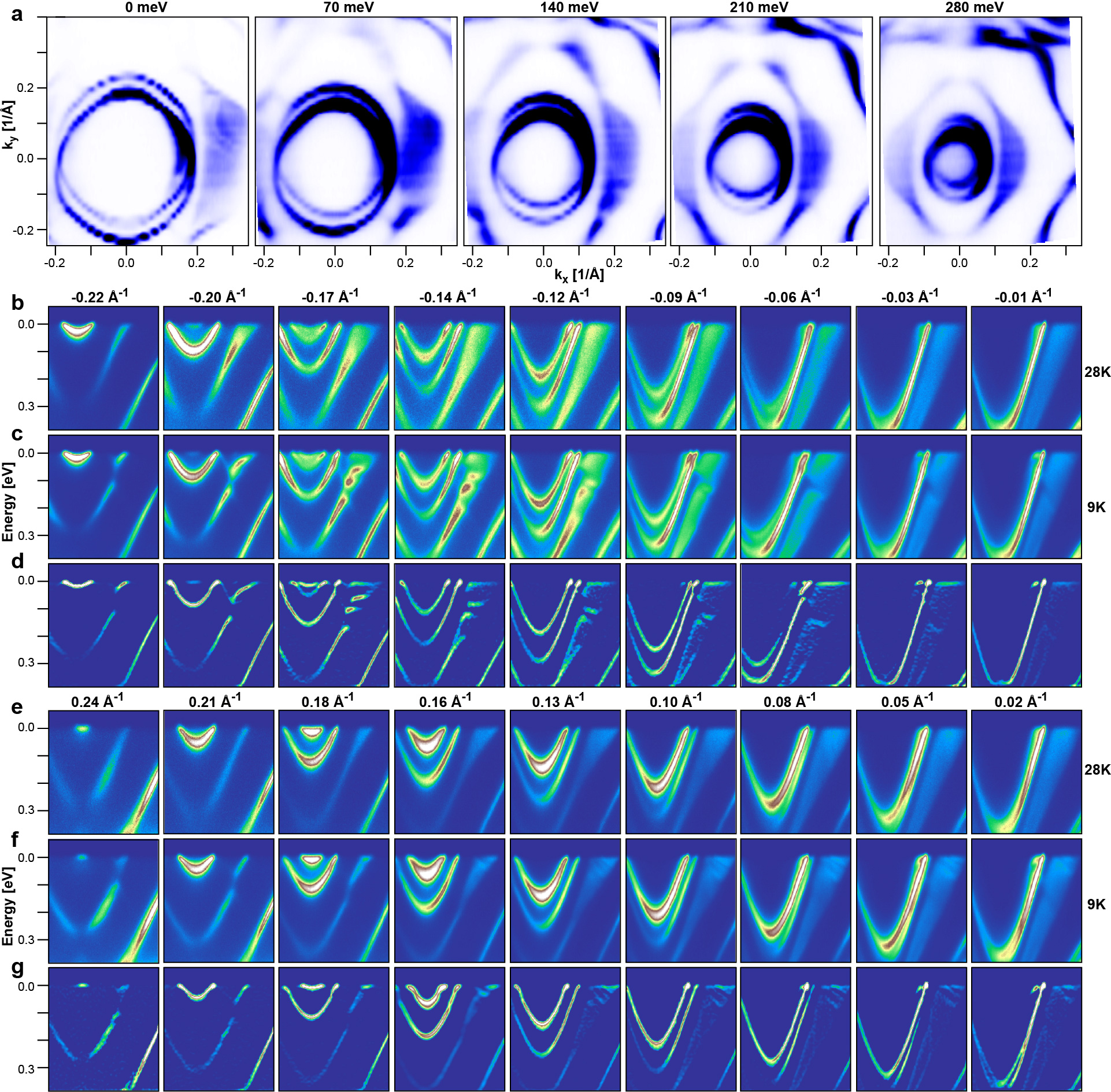}%
    \caption{Additional data from the dataset presented in Fig.~4a-c.
    (a) Fermi surface map and equal-energy maps measured at T~=~9~K.
    (b,~e) and (c,~f) Spectra measured parallel to the $\Gamma-K$ direction at different k$_y$s at T~=~28 and 9~K, respectively.
    (d,~g) curvature plots \cite{zhang2011precise} for the spectra in (c,~f).    
    }
\end{figure*}

\section{Tb~$5d$ band: temperature dependence.}
In Fig.~7, we demonstrate the temperature evolution of the band structure along a cut taken through the Tb~$5d$ band (the part where the gap opens). This is a dataset from which the EDCs in Fig.~4d were extracted. The folded band and the hybridization gap can be identified in the spectra measured at temperatures up to 19~K. At T~=~20~K, the folded band cannot be identified, but the hybridization gap is still present. At T~=~21~K, the gap disappears, and upon further temperature increase, the bands do not change significantly. This result agrees with the early reported transition temperature of 20.4~K \cite{guo20241, ortiz2024intricate}.
\begin{figure}[b]
    \includegraphics[width=1\linewidth]{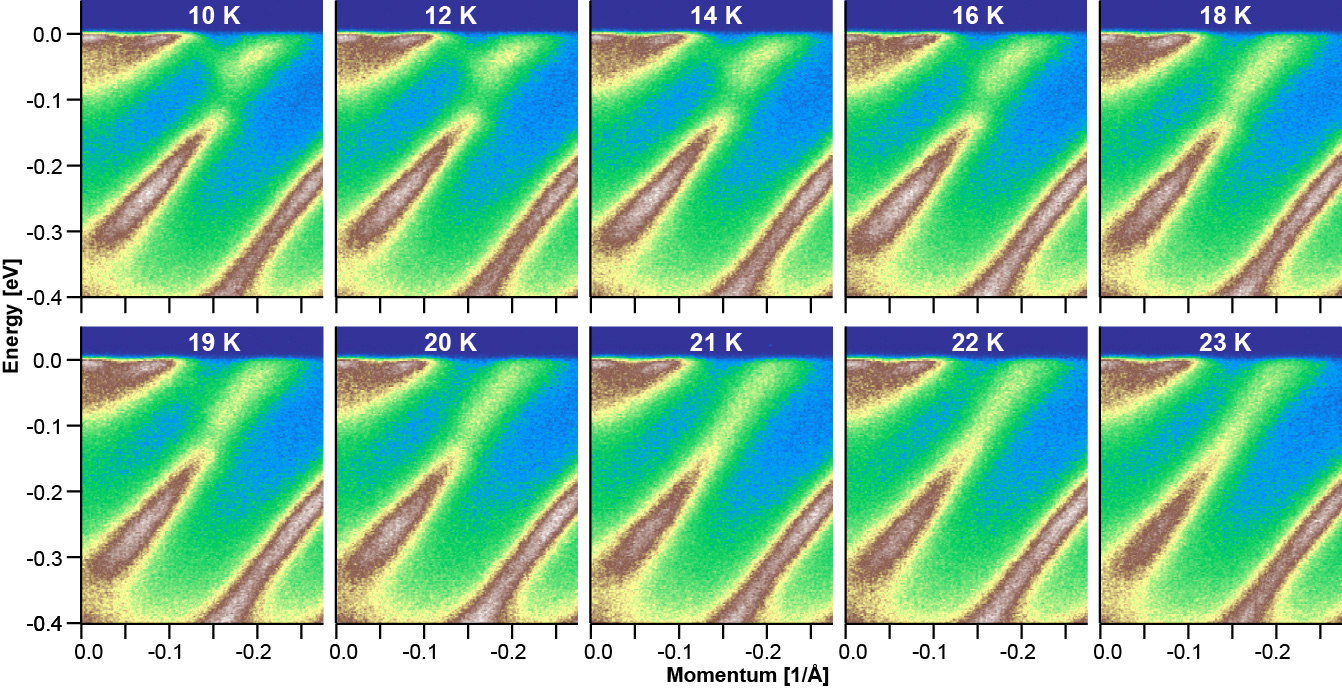}%
    \caption{Spectra measured parallel to the $\Gamma-K$ direction at k$_y$=~0.195~\AA$^{-1}$ at different temperatures from 10 to 23~K.  
    }
\end{figure}

\section{Splitting of the electron-like band: alternative interpretation.}
The alternative interpretation of the Fermi surface pockets observed around the $\Gamma-$point would be two ellipses with the same size along $\Gamma-K$ but different sizes along $\Gamma-M$. Although we cannot rule out this scenario completely without a proper theoretical explanation of the observed splitting, our experimental data speaks in favor of the two identical, shifted circles interpretation. From the Fermi surface in Fig.~4a, one can see that the intensity of the inner contour in the top-left quadrant is high, as is the intensity of the outer contour in the bottom-left quadrant. Together, they form one continuous circular contour without any significant jumps in intensity. At k$_y$~=~0, this contour crosses a significantly less intense contour formed by the outer contour in the top-left quadrant and the inner contour in the bottom-left quadrant. A similar pattern can also be seen in other equal-energy contours shown in Fig.~6a.

\section{DFT culculations}
In Fig.~8, we show the DFT-calculated Fermi surfaces and high-symmetry cuts for Tb-terminated TbTi$_3$Bi$_4$.

\begin{figure*}[b]
    \includegraphics[width=0.8\linewidth]{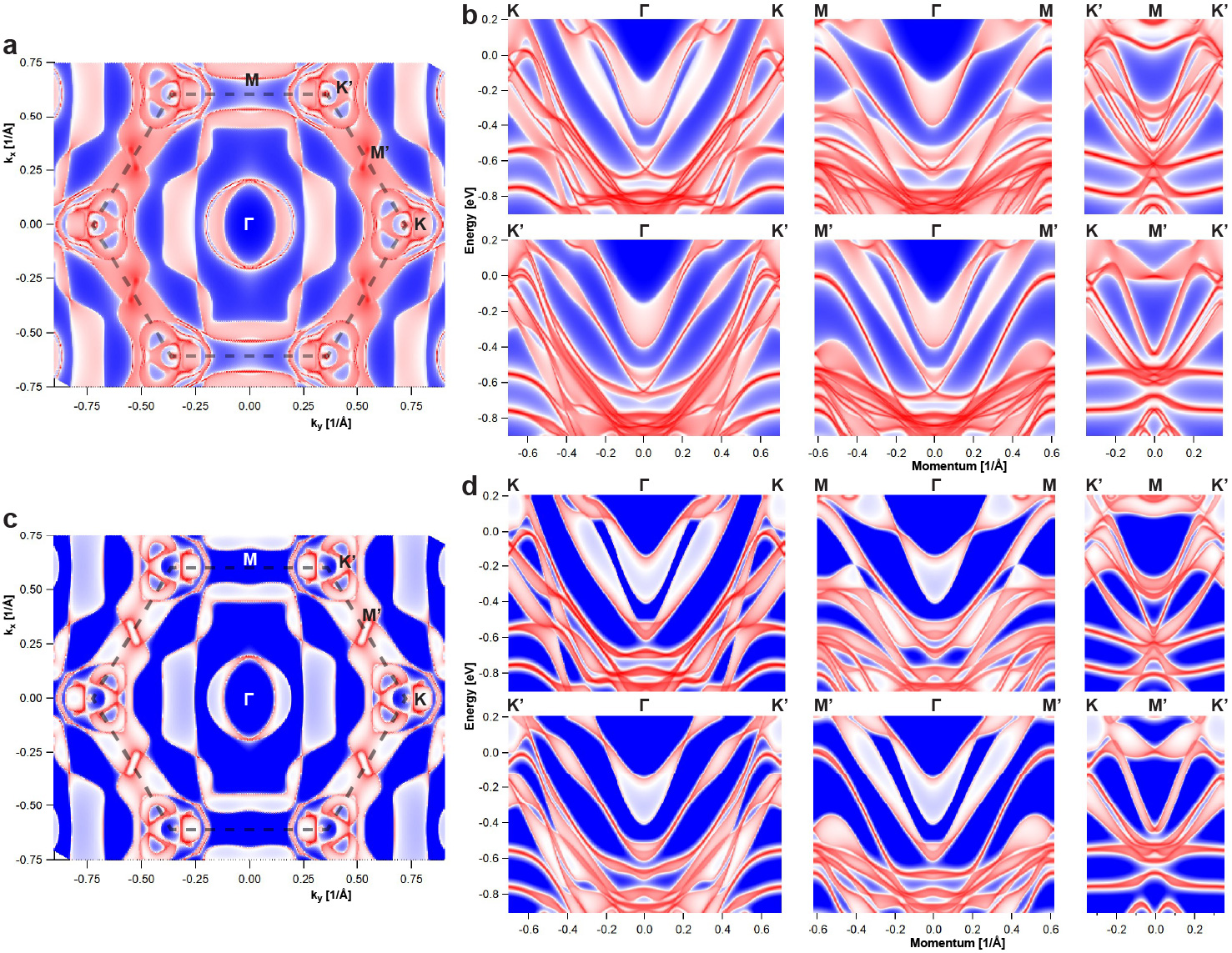}%
    \caption{DFT calculations.
    (a) Fermi surface map from semi-infinite (001) surface calculations.
    (b) High-symmetry cuts from semi-infinite surface calculations.
    (c,~d) The same as (a,~b) but only bulk states.
    }
\end{figure*}

\section{X-ray diffraction.}
The sample is malleable and does not grind well, smearing and deforming too much for powder x-ray diffraction measurements. Hence, diffraction measurements were carried out on single-crystal samples, which were cleaved along the (00l) plane, using a Rigaku MiniFlex~II powder diffractometer in Bragg-Brentano geometry with Cu~K$\alpha$ radiation ($\lambda$~=~1.5406~\AA), following the method described in Ref.~\cite{jesche2016x}. The results are shown in Fig.~9 and confirm the TbTi3Bi4 phase. The obtained lattice parameter c~=~24.743(3)~\AA~is comparable to the reported value c~=~24.785(4)~\AA~obtained using Mo~K$\alpha$ radiation ($\lambda$~=~0.71073~\AA) in Ref.~\cite{guo20241}.

\begin{figure}[b]
    \includegraphics[width=0.91\linewidth]{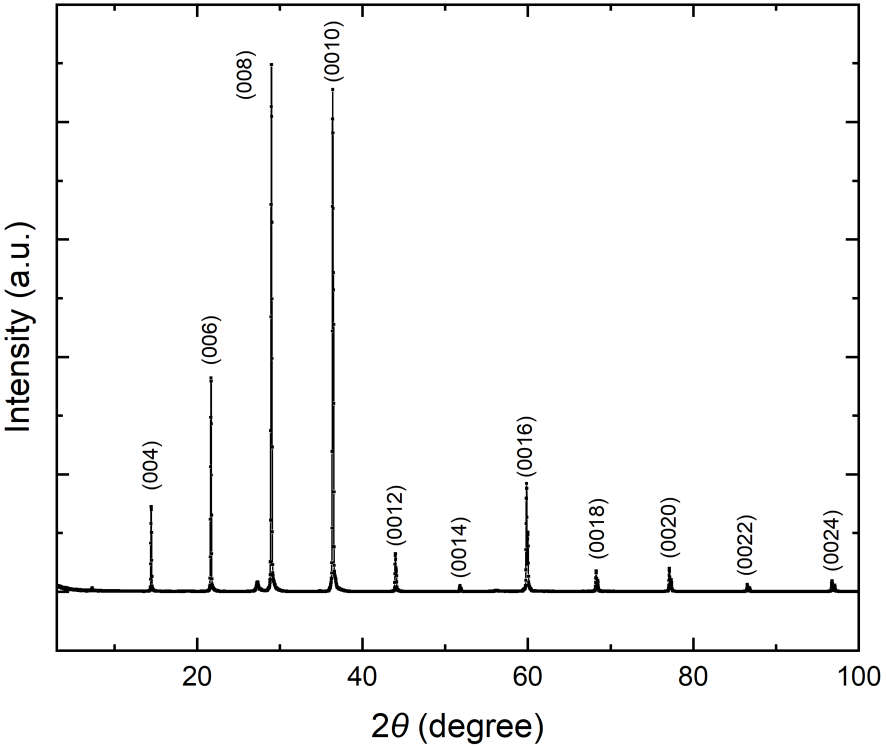}%
    \caption{The results of single-crystal X-ray diffraction measurement along the (00l) direction.
    }
\end{figure}

\bibliography{ndBi_arcs}

\end{document}